\documentclass[amsmath,amssymb,pra,aps,showpacs,superscriptaddress,twocolumn]{revtex4-2}
\usepackage{amsmath,amsfonts,amssymb,amsthm,graphics,graphicx,epsfig,bbm}
\usepackage[colorlinks=true,citecolor=blue,linkcolor=blue,urlcolor=blue]{hyperref}
\usepackage[usenames]{color}
\usepackage{graphicx}
\usepackage{subfigure}
\usepackage{amsmath}
\usepackage{epsfig}
\usepackage{dcolumn}
\usepackage{bm}
\usepackage{color}
\usepackage{times}
\usepackage{epstopdf}
\usepackage{amssymb}
\usepackage{amstext}
\usepackage{latexsym}
\usepackage{hyperref}
\usepackage{amsfonts}
\usepackage{psfrag}
\usepackage{soul,xcolor}
\usepackage[normalem]{ulem}
\usepackage{dsfont}
\usepackage{txfonts}
\usepackage{float}

\newcommand{\ket}[1]{\vert #1 \rangle}
\newcommand{\bra}[1]{\langle #1 \vert}

\begin{document}

\setstcolor{red}

\title{Quantum simulation of entanglement dynamics in a quantum processor} 
\date{\today}

\author{Sebasti\'an Saavedra-Pino}
\affiliation{Departamento de Física, Universidad de Santiago de Chile (USACH), Avenida Víctor Jara 3493, 9170124, Santiago, Chile}

\author{Cristian Inzulza}
\affiliation{Departamento de Ingenier\'ia Inform\'atica, Universidad de Santiago de Chile (USACH), Santiago, Chile}

\author{Pablo Rom\'an}
\affiliation{Departamento de Ingenier\'ia Inform\'atica, Universidad de Santiago de Chile (USACH), Santiago, Chile}

\author{Francisco Albarr\'an-Arriagada\footnote{Corresponding author}}
\affiliation{Departamento de Física and Center for the Development of Nanoscience and Nanotechnology, Universidad de Santiago de Chile (USACH), Avenida Víctor Jara 3493, 9170124, Santiago, Chile}
\email[F. Albarr\'an-Arriagada]{\qquad francisco.albarran@usach.cl}

\author{Juan Carlos Retamal}
\affiliation{Departamento de Física and Center for the Development of Nanoscience and Nanotechnology, Universidad de Santiago de Chile (USACH), Avenida Víctor Jara 3493, 9170124, Santiago, Chile}

\begin{abstract}
We implement a five-qubit protocol in  IBM quantum processors to study entanglement dynamics in a two qubit system in the presence of a simulated environment.  Specifically,  two qubits represent the main system, while another two qubits serve as the environment. Additionally, we employ an auxiliary qubit to estimate the quantum entanglement. Specifically, we observe the sudden death and sudden birth of entanglement for different inital conditions that were simultaneously implemented on the IBM 127-qubit quantum processor \textit{ibm$\_$brisbane}. We obtain the quantum entanglement evolution of the main system qubits and the environment qubits averaging over $N=10$ independent experiments in the same quantum device. Our experimental data shows the entanglement and disentanglement signatures in system and enviroment qubits, where the noisy nature of current quantum processors produce a shift on times signaling sudden death or sudden birth of entanglement. This work takes relevance showing the usefulness of current noisy quantum devices to test fundamental concepts in quantum information.
\end{abstract}

\maketitle

\section{Introduction}
During the last decades, quantum computing has experienced remarkable growth, largely due to the impressive experimental development, which has allowed the rising of different quantum platforms for the implementation of noisy intermediate-scale quantum (NISQ) processors to test quantum protocols~\cite{Bharti2022RevModPhys, Preskill2018Quantum}. Some of the most remarkable platforms for such NISQ devices are quantum dots~\cite{Zhang2019NatlSciRev, Loss1998PhysRevA}, neutral atoms~\cite{Levine2019PhysRevLett, Henriet2020Quantum}, trapped ions~\cite{Bruzewicz2019ApplPhysRev}, photonic circuits~\cite{Pelucchi2021NatRevPhys}, and superconducting circuits~\cite{Huang2020SciChinaInfSci, Kjaergaard2020AnnuRevCondensMatterPhys}, where the last two have reported quantum advantage experiments, that is, an algorithmic task that can be performed by a quantum computer several times faster than a classical one \cite{Zhong2021PhysRevLett, Madsen2022Nature, Wu2021PhysRevLett, Morvan2023arXiv}. Even if quantum computers have experimented a fast enhancement in the coherence time of the qubits and connectivity among them, it is still far away from the fault-tolerant quantum computer~\cite{Fowler2012PhysRevA}. 

On the other hand, quantum simulation and experimental test of  fundamental effects in quantum mechanics and condensed matter physics is a good scenario to obtain some practical advantage in the current NISQ era~\cite{Daley2022Nature}. In this line, the remote use of quantum devices through cloud services like IBM quantum cloud~\cite{IBMQ} provide us an accesible tool to test quantum protocols experimentally either for algorithmic applications with few qubits with the potential to present some advantage for near future, or fundamental phenomenology hard to test in natural systems. 

One of the most interesting  concepts in quantum mechanics is quantum entanglement, playing a central role  in quantum information processing and quantum computing, as a resource that enables powerful computations and secure communication protocols~\cite{Horodecki2009RevModPhys}. A central issue concerning entanglement is the estimation of the amount of entanglement embedded in a bipartite quantum system. For two qubits, this problem is well understood and we can estimate entanglement using as a figure of merit, the Concurrence introduced by Wooters in 1998 \cite{wooters1998}. How is the entanglement affected by the environment is an issue widely explored in quantum information. In particular describing entanglement dynamics  in the presence of environment for different initial conditions among a system of two qubits, lead to  the emergence of the so called entanglement sudden death (ESD) effect~\cite{eberly2004}. Conversely focusing on entanglement among reservoirs, lead to the emergence of sudden birth of entanglement (ESB) \cite{Lopez2008PhysRevLett}. This is an issue that has been experimentally addressed using propagating photons \cite{davidovich1, davidovich2, davidovich3}.

To study the entanglement  dynamics take relevance in the context of superconducting quantum processor, since they are  suitable for the experimental simulation of open quantum systems~\cite{GarciaPerez2020NpjQuantumInf}.  This is the aim of the present work, estimate the entanglement among qubits interacting with an environment, using the  \textit{ibm$\_$brisbane}  quantum processor. For certain initial conditions the concurrence can be estimated  measuring an entanglement witness, originally proposed to be implemented in cavity QED experiment \cite{Santos2006PhysRevA}. We implement a protocol to simulate the dynamics of two qubits each one affected by and independent environment, modeled by two additional qubits. Although, the ESD could appears in different scenarios for open quantum systems, here we consider specifically, the amplitud damping case as is considered in Ref.~\cite{Lopez2008PhysRevLett}. Aditionally, we estimate the concurrence by the measuring of an entanglement witness with the aid of an auxiliary qubit. As we will see the experimental measurement, as compared with the ideal entanglement dynamics, experience a deviation due to the intrinsic noise device, shifting characteristic times for ESD and ESB. 

 \begin{figure}[t]
	\centering
	\includegraphics[width=0.8\linewidth]{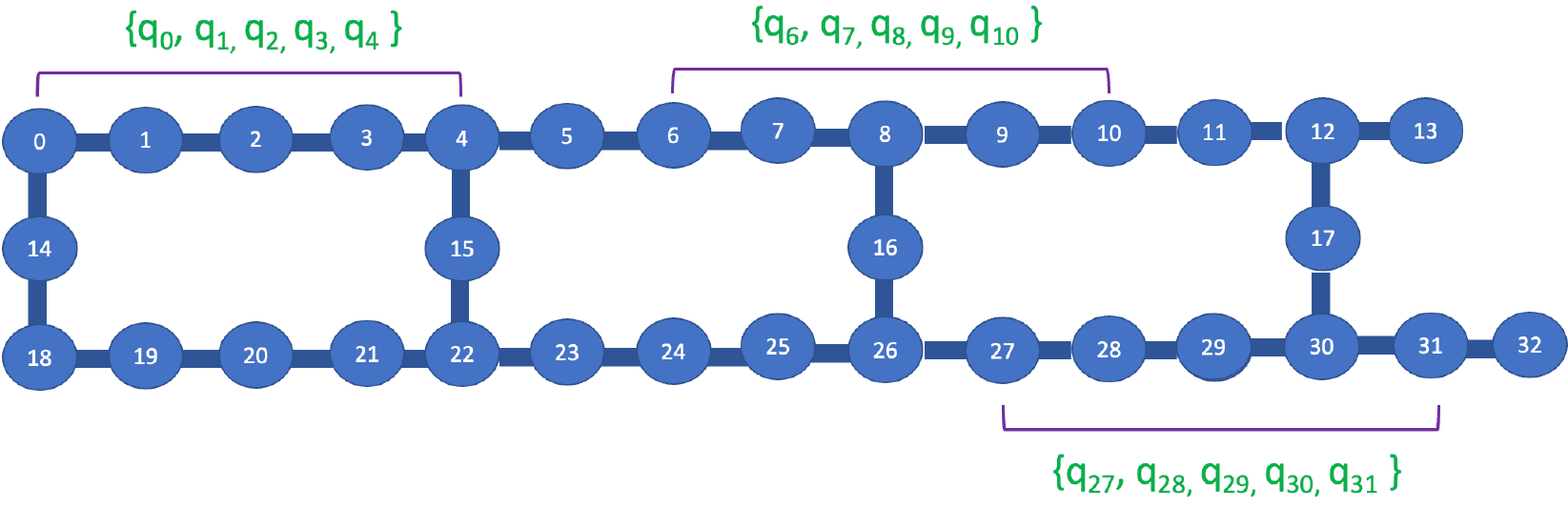}
	\caption{Connectivity between qubits in \textit{ibm$\_$brisbane}  quantum processor.} 
	\label{Fig01}
\end{figure}

\section{Quantum circuit implementation}\label{sec2}

To measure the entanglenment among two qubits, we use five qubits, two of them are used as the main system, the other two represent the reservoir (one per main system qubit), and an extra one is used to estimate the entanglement between the main system qubits and between the reservoir qubits. 
 We divide the circuit implementation into three well-defined steps. First, the initial quantum state preparation; second, the system-environment evolution; and third, the entanglement estimation process. As we mentioned before, we use the 127-qubit quantum procesor \textit{ibm$\_$brisbane}. In Figure 1 we show the connectivity for the first 32 qubits of the 127 qubits of this processor. We plan to implement three experiments as we will see later, each one considering five qubits. Brisbane architecture allow us to carry out three independent simultaneous experiments,  choosing three independent sets of five qubits with linear connectivity. For example, the set one $\{q_0, q_1,q_2, q_3, q_4\}$, set two $\{q_6, q_7,q_8, q_9, q_{10}\}$ and set three $\{q_{27}, q_{28},q_{29}, q_{30}, q_{31}\}$.  On each one of these three sets, we encode a state $|\psi\rangle=\alpha |00\rangle+\beta|11\rangle$ for corresponding amplitudes $\alpha = 1/\sqrt{2},1/\sqrt{3},1/\sqrt{5}$. In addition we will consider permutations of the encoding to test every case on each set of qubits. 
 
 To illustrate the protocol let us consider the set one. In this case we use qubits labeled as $q_1$ and $q_3$ as the system qubits, as environment the qubits $q_0$ and $q_4$, and as the auxiliary qubit to measure the entanglement we use the qubit $q_2$. Before starting to discuss the different steps in the quantum circuit, we mention that the \textit{ibm$\_$brisbane}  quantum processor has a set of basic universal gates, and any other gate can be  decomposed as a combination of them. The set of basic gates are the nearest neighbors echoed cross-resonance gate (ECR), identity gate (ID), $z-$rotation ($RZ$), bit flip or $X$ gate, and $\sqrt{X}$ that has a matrix representation given by:
\begin{eqnarray}
\sqrt{X}=\frac{1}{2}
\begin{pmatrix}
1+i & 1-i \\
1-i & 1+i
\end{pmatrix}.
\label{Eq01}
\end{eqnarray}

\subsection{Initial state preparation}\label{subsec2}

\begin{figure}[t]
	\centering
	\includegraphics[width=0.5\linewidth]{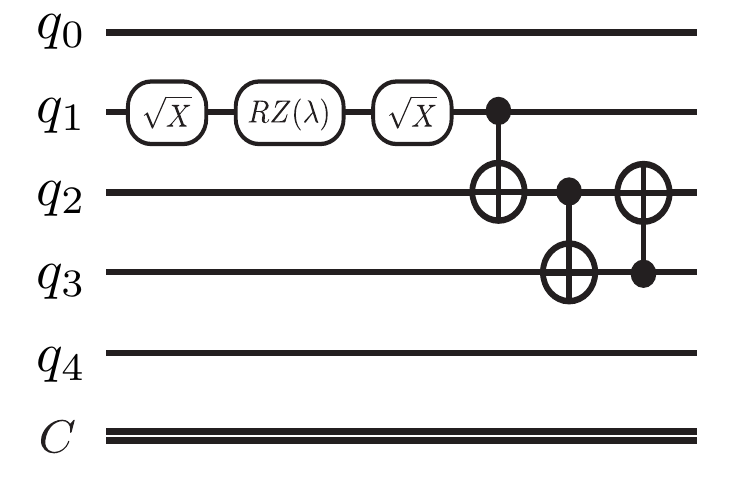}
	\caption{Quantum circuit for the initial state preparation given by Eq.~(\ref{Eq02}).} 
	\label{Fig02}
\end{figure}

\begin{figure}[t]
	\centering
	\includegraphics[width=0.7\linewidth]{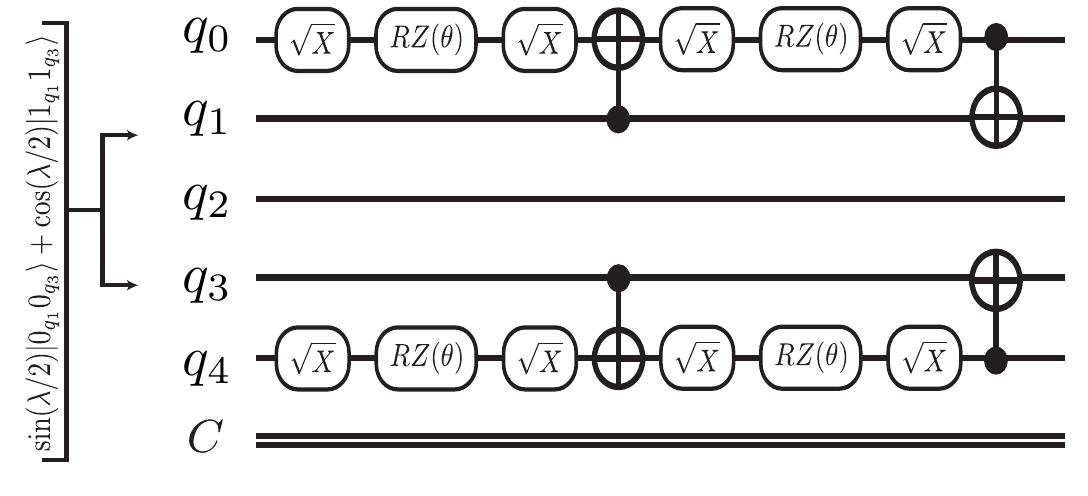}
	\caption{Quantum circuit for the system-environment evolution given by Eq.~(\ref{Eq05})}
	\label{Fig03}
\end{figure}

\begin{figure}[t]
	\centering
\includegraphics[width=1\linewidth]{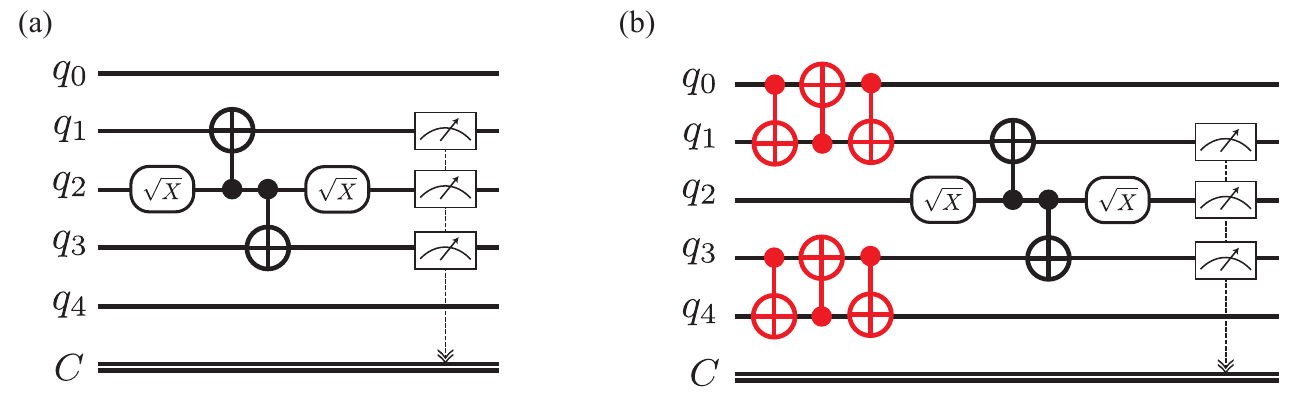}
	\caption{Quantum circuit for entanglement estimation for system qubits (a) and environment qubits (b) shown in Sec.~\ref{subsec2}} 
	\label{Fig04}
\end{figure}

We will focus on the preparation of entangled states of the form
\begin{equation}
\ket{\psi}=\sin(\lambda/2)\ket{0_{q_1}0_{q_3}} + \cos(\lambda/2)\ket{1_{q_1}1_{q_3}}.
\label{Eq02}
\end{equation}
To do this, we perform a rotation between the state $\ket{0}$ and $\ket{1}$ of qubit $q_1$ and then a control-not gate  using $q_1$ as control and $q_3$ as the target. For the single qubit rotation we can use the sequence $\sqrt{X}\,RZ(\lambda)\,\sqrt{X}$ obtaining the transformation
\begin{eqnarray}
\sqrt{X}\,RZ(\lambda)\,\sqrt{X}=
\begin{pmatrix}
\sin(\lambda/2) & \cos(\lambda/2) \\
\cos(\lambda/2) & -\sin(\lambda/2) 
\end{pmatrix}.
\label{Eq03}
\end{eqnarray}

Now, the control-not gate from $q_1$ to $q_3$ cannot be done directly due to the connectivity of the chip. To perform it, first, we do a CN gate from $q_1$ to $q_2$, a CN gate from $q_2$ to $q_3$ and another from $q_3$ to $q_2$. This gate scheme is shown in Fig.~\ref{Fig02}, where the last two gates do the swapping between $q_2$ and $q_3$.

\subsection{System-environment evolution}\label{subsec2}

Now, we consider that each system qubit evolves under an amplitude-damping channel that is described by the master equation
\begin{equation}
\dot{\rho}_S=\gamma\sum_{j=\{1,3\}}\left[\sigma_{-,j}\rho_S\sigma_{+,j} - \frac{1}{2}\{\sigma_{+,j}\sigma_{-,j},\rho_S\}\right],
\label{Eq04}
\end{equation}
where $\rho_S$ is the density matrix that describes the state of the system qubits ($q_1$ and $q_3$), $\gamma$ is the decay rate, and $\sigma_{\pm,j}$ are the raising and lower operator for the qubit $q_j$. Each qubit decays independently under its own reservoir. After some calculations it can be shown that  the dynamics for each qubit can be described by the dissipative map
\begin{eqnarray}
&&\ket{0_{q_{1(3)}}0_{q_{0(4)}}} \rightarrow \ket{0_{q_{1(3)}}0_{q_{0(4)}}}\nonumber\\
&&\ket{1_{q_{1(3)}}0_{q_{0(4)}}} \rightarrow \eta\ket{1_{q_{1(3)}}0_{q_{0(4)}}} + \zeta\ket{0_{q_{1(3)}}1_{q_{0(4)}}}\quad
\label{Eq05}
\end{eqnarray}
with $\eta=e^{-\gamma t/2}$, $\zeta=\sqrt{1-e^{-\gamma t}}$, and $q_{0(4)}$ is the reservoir degree of freedom (purification space) corresponding to the system qubit $q_{1(3)}$. Under this map, we can obtain the next density matrix of the system qubits $\rho_S(t)$ (after tracing the reservoir degrees of freedom)

\begin{eqnarray}
\rho_S(t)=\begin{pmatrix}
s^2 + c^2\eta^4& 0 & 0 & sc\eta^2 \\
0 & c^2\eta^2\zeta^2& 0 & 0 \\
0& 0 & c^2\eta^2\zeta^2 & 0 \\
sc\eta^2& 0 & 0 & c^2\zeta^4 
\end{pmatrix},
\label{Eq06}
\end{eqnarray}
where $c=\cos(\lambda/2)$ and $s=\sin(\lambda/2)$ are the initial condition given in Eq.~(\ref{Eq02}). The circuit implementation of this map is given in Fig.~\ref{Fig03}, with $\theta=\sin^{-1}(e^{-\gamma t/2})$. In ref.~\cite{GarciaPerez2020NpjQuantumInf}, the authors show how to implement different maps for open quantum systems in IBM quantum processors beyond the amplitude damping. Particularly they show how to use the IBM quantum computer as a testbed for Markovian and non-Markovian dynamics in open quantum systems.

\subsection{Entanglement estimation}\label{subsec2}

Finally, the central issue of this work is the estimation of  entanglement. To do this, we will estimate the concurrence of the system qubits state, and environment qubits state following the proposal described in ref.~\cite{Santos2006PhysRevA}. As the density matrix of the qubit systems given in Eq.~(\ref{Eq06}) is an $X$-form matrix, then, the concurrence for the system qubits $\mathcal{C}_S$ is given by
\begin{eqnarray}
\mathcal{C}_S=\max\left[0,-2\eta^2(c^2\zeta^2-sc)\right].
\label{Eq07}
\end{eqnarray}

We note that $\eta^2sc$ is the matrix element $\bra{00}\rho_S\ket{11}$. As $\eta^2+\zeta^2=1$, we have that $\eta^4=\eta^2(1-\zeta^2)$ and $\zeta^4=\zeta^2(1-\eta^2)$, obtaining $\bra{00}\rho_S\ket{00}=s^2+c^2\eta^2-c^2\eta^2\zeta^2$ and $\bra{11}\rho_S\ket{11}=c^2\zeta^2-c^2\eta^2\zeta^2$. Therefore, the probability to obtain the entangled state $\ket{\Phi}=\sqrt{\frac{1}{2}}(\ket{00}+\ket{11})$ is related to the concurrence as
\begin{eqnarray}
&&P_{\Phi}=\bra{\Phi}\rho_S\ket{\Phi}=\frac{1}{2}\left[1-2\eta^2(c^2\zeta^2-cs)\right]\nonumber\\
\Rightarrow && \mathcal{C}_S=\max \left\{0,2P_{\Phi}-1\right\}.
\label{Eq08}
\end{eqnarray}
Now, we note that the state $\ket{\Phi}$ can be obtained from $\ket{00}$ as
\begin{eqnarray}
\ket{\Phi}=\sqrt{\frac{1}{2}}(\mathbb{I}+\sigma_x\otimes\sigma_x)\ket{00}.
\label{Eq09}
\end{eqnarray}

The transformation given by Eq.~(\ref{Eq09}) can be implemented using the circuit sketched in Fig.~\ref{Fig04} a. (Fig.~\ref{Fig04} b. for the environment qubits). If we consider that the state of the quantum processor after the evolution given by Fig.~\ref{Fig03} is given by $\ket{\xi}\ket{0_{q_2}}$, where $\ket{\xi}$ is the state of the qubits $q_1$, $q_3$, $q_4$, and $q_4$; the state after evolution under the circuit given by Fig.~\ref{Fig04} a. (before the measurement process) reads
\begin{equation}
\frac{1}{2}\left[i(\ket{\xi}-\sigma_x^{q_1}\sigma_x^{q_3}\ket{\xi})\ket{0_{q_2}} + (\ket{\xi}+\sigma_x^{q_1}\sigma_x^{q_3}\ket{\xi})\ket{1_{q_2}}\right],
\label{Eq10}
\end{equation}
then, the probability $\mathcal{P}_{010}$ to measure $0_{q_1}1_{q_2}0_{q_3}$ is given by
\begin{equation}
\mathcal{P}_{010}=\frac{P_{\Phi}}{2}.
\label{Eq11}
\end{equation}
Thus we can relate the concurrence with the probability to measure the qubits $q_1q_2q_3$ in $|010 \rangle$ leading to $\mathcal{C}_S=\max\left\{0,4\mathcal{P}_{010}-1\right\}$. 

Finally, to get the entanglement dynamics for the environment qubits, we note that the reduced density matrix that described them after the system-environment step (analog to Eq.~\ref{Eq06}, but now for the environment qubits) also has $X$-form similar to Eq.~\ref{Eq06}, then the concurrence can be calculated using the same protocol. Nevertheless, as the environment qubits $q_0$ and $q_4$ are not directly connected to the auxiliary qubit $q_2$ we need to perform a swap gate between the qubit $q_0$ and $q_1$, as well as between the qubits $q_3$ and $q_4$, and repeated the previous protocol as is shown in Fig.~\ref{Fig04} b. In that figure, we can observe that the first CN gates (in red) perform the mentioned swap transformation. Also, we note that the measurement process is again over the qubits $q_1$, $q_2$ and $q_3$, avoiding extra swap gates needed if we want to measure the qubits $q_0$ and $q_4$, initially labeled as environment qubits, instead $q_1$ and $q_3$.

\section{Results}\label{sec3}

As we have previously stated, we have implemented our protocol in \textit{ibm$\_$brisbane}  quantum processor with connectivity shown in Fig.~\ref{Fig01}.  We consider three cases for the initial entangled pair of system qubits $|\psi\rangle =\alpha |00\rangle+\beta |11\rangle$, to compare with analytical results in Ref. \cite{Lopez2008PhysRevLett}. The first case for  $\alpha =\sqrt{1/2}$ where correlations in the system qubits decay asymptotically, whereas correlations in the reservoir qubits growth starting in $t=0$.  The second case for unbalanced populations in the initial state with $\alpha=\sqrt{1/3}$ where correlations in the system qubits suddenly disappear in a time $t_d$, as well as correlations in reservoir qubits suddenly appear for a time $t_b < t_d$. The third  case for $\alpha =\sqrt{1/5}$ for which correlations in the system qubits suddenly disappear in a time $t_d$, and correlations in reservoir qubits appear abruptly at the same time $t_b=t_d$. 

\begin{figure}[h]
	\centering
	\includegraphics[width=0.8\linewidth]{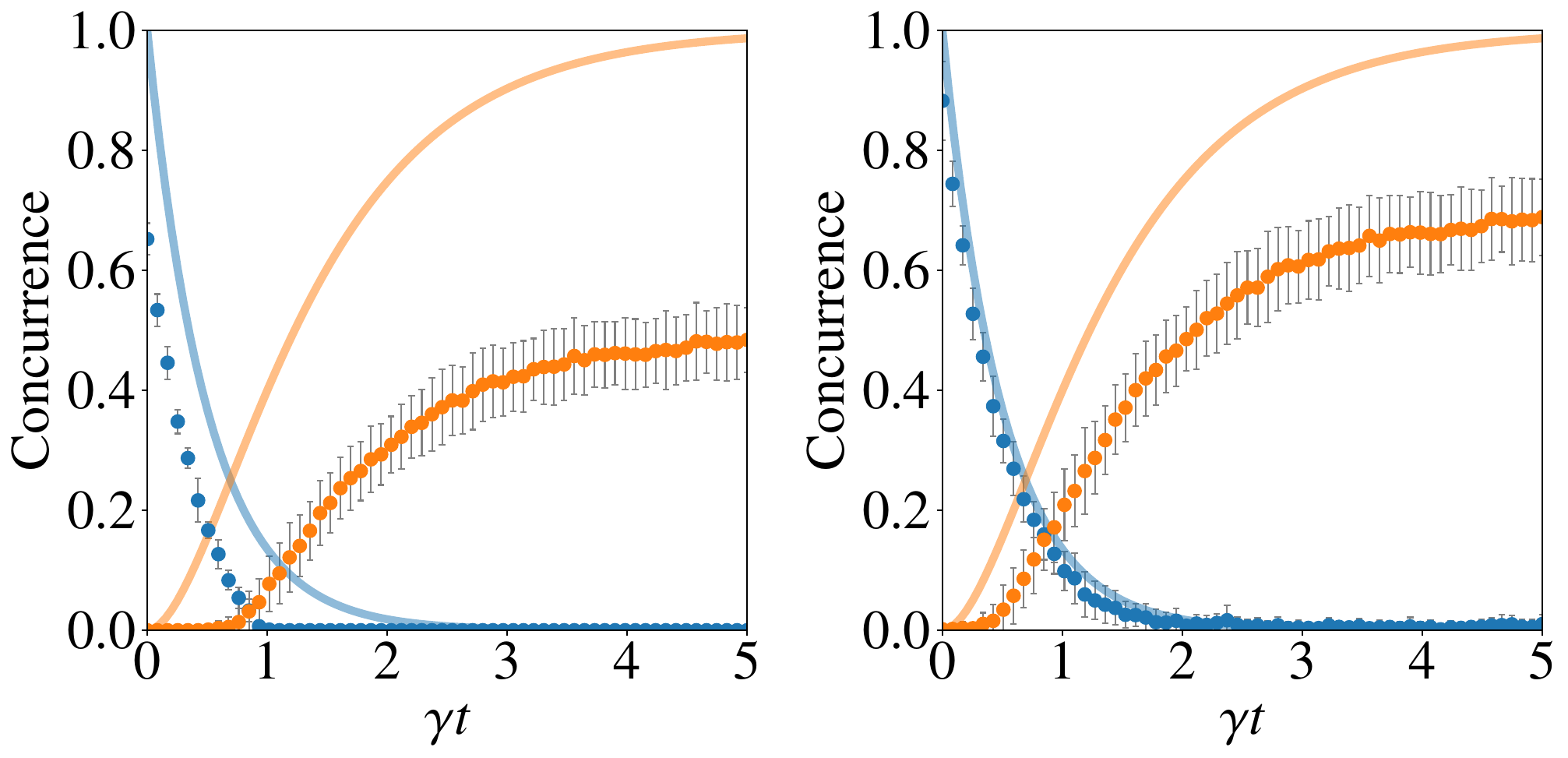}
	\caption{Concurrence as a function of $\gamma t$ for the the amplitude $\alpha =1/\sqrt{2}$ obtained from \textit{ibm$\_$brisbane}  for the case of  without (left panel) and with error mitigation (right panel). Orange dots are for environment concurrence and blue dots for system concurrence. The solid curves show the theoretical results.}
	\label{Fig05}
\end{figure}

\begin{figure}[h]
	\centering
	\includegraphics[width=0.8\linewidth]{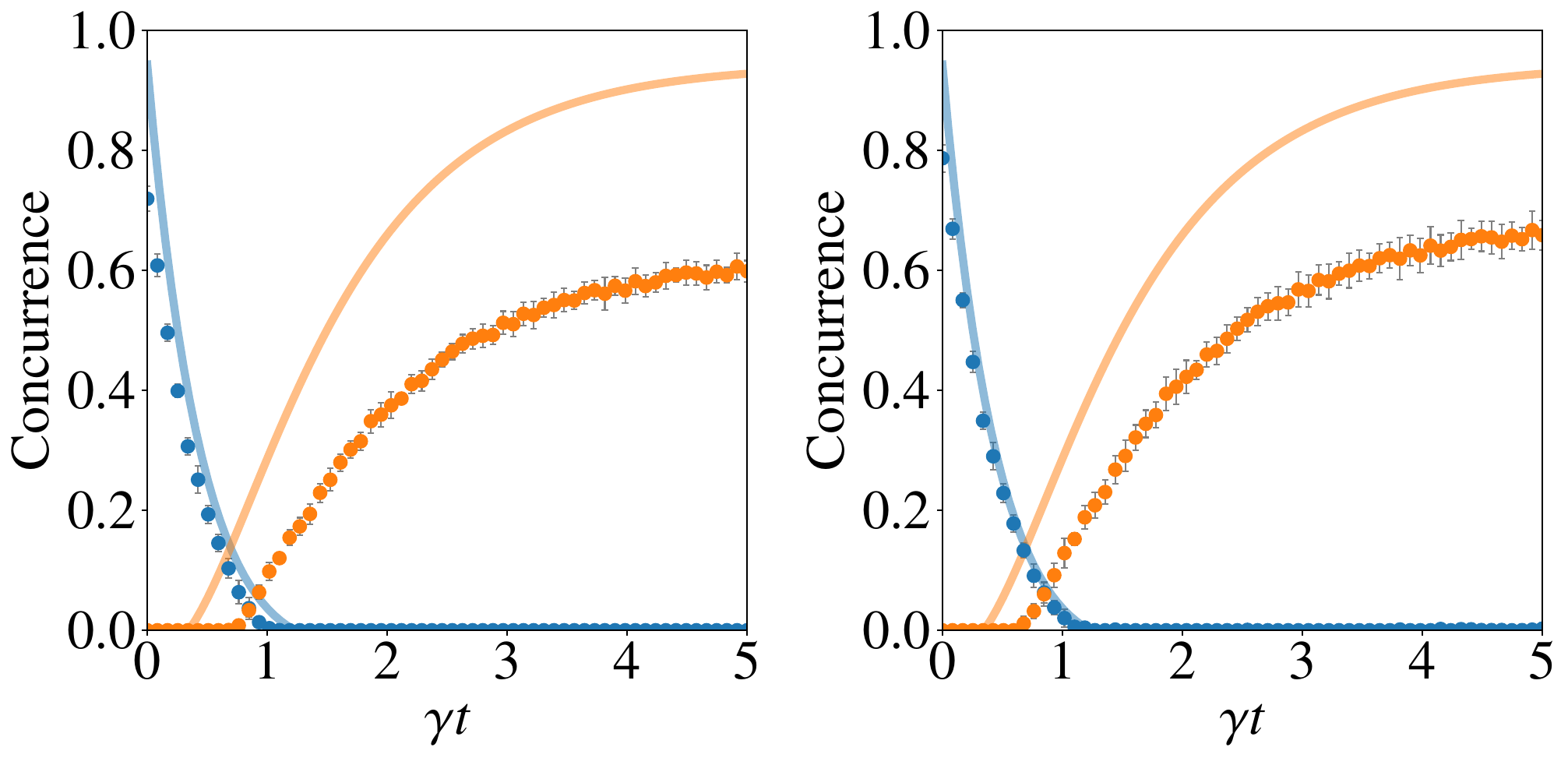}
	\caption{Concurrence as a function of $\gamma t$ for the the amplitude $\alpha =1/\sqrt{3}$ obtained from \textit{ibm$\_$brisbane}  for the case of  without (left panel) and with error mitigation (right panel). Orange dots are for environment concurrence and blue dots for system concurrence. The solid curves show the theoretical results.}
	\label{Fig06}
\end{figure}

\begin{figure}[h]
	\centering
	\includegraphics[width=0.8\linewidth]{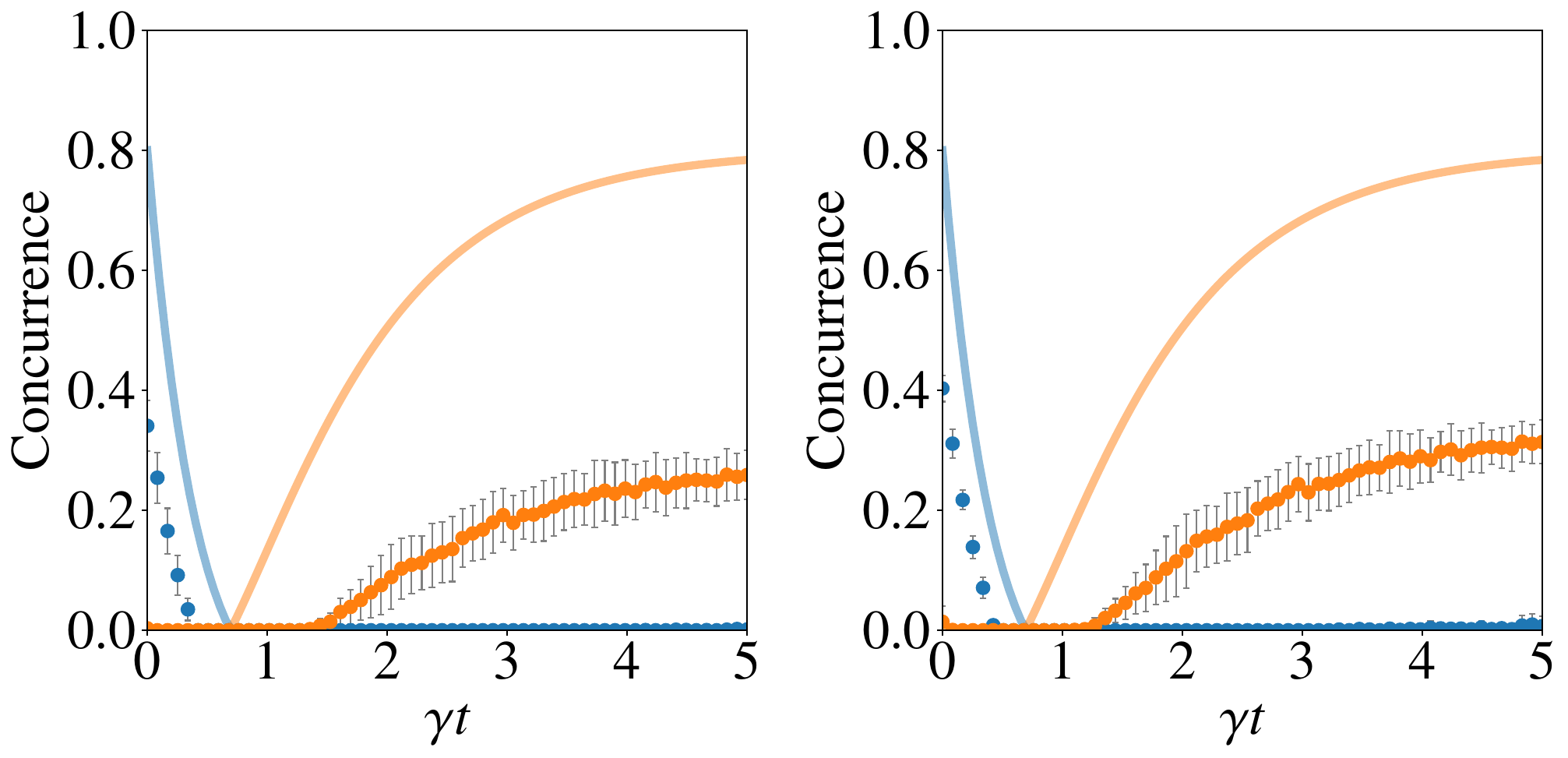}
	\caption{Concurrence as a function of $\gamma t$ for the the amplitude $\alpha =1/\sqrt{5}$ obtained from \textit{ibm$\_$brisbane}  for the case of  without (left panel) and with error mitigation  (right panel). Orange dots are for environment concurrence and blue dots for system concurrence. The solid curves show the theoretical results.}
	\label{Fig07}
\end{figure}

The measurement carried out in \textit{ibm$\_$brisbane} is shown in Fig. ~\ref{Fig05}-\ref{Fig07}, for three cases considered, $\alpha=1/ \sqrt{2}, \alpha=1/ \sqrt{3},\alpha=1/ \sqrt{5}$ wich are implemented simultaneously, in one experiment, in set 1, set 2 and set 3 respectively. We run  $N=10$ experiments, each experiment consisting on 20000 shots. Fig. ~\ref{Fig05}-\ref{Fig07} shows the average value for the entanglement estimator for each case over these $N=10$ experiments. The corresponding errors for the average values are indicated by error bars. Left panel shows the entanglement estimator for system and environment not including error mitigation techniques, and right panel including error mitigation~\cite{PaulDNation2021PRXQuantum}. The effect of the noise device is evident in these results, when we compare with the theoretical prediction for these cases \cite{Lopez2008PhysRevLett}, shown in each figure with solid lines. The effect of  the  noise device is superposed with  the decay  modeled through the different steps of the protocole (Fig. \ref{Fig01}-\ref{Fig03}) leading to a shift of times signaling the sudden death or the sudden birth point. Also noise is affecting the entanglement amplitude as compared with theoretical prediction. 

In addition, we observe that the three sets of qubits do not have the same performance. The set three lead us with an entanglement estimation that deviates substantially from the theoretical case. The set 1 and set 2, give results which compares favorable with theoretical results. This is shown in appendixes A and B, that contain the experiments for $\alpha=1/ \sqrt{2}
, \alpha=1/ \sqrt{3},\alpha=1/ \sqrt{5}$ in set 2,3, and 1 and set 3,1 and 2 respectively. It is important to mention that the use of different sections of the quantum chip to perform parallel experiments is an interesting application of large qubits quantum processor. Nevertheless, as the physical properties change from qubit to qubit, such as relaxation times ($T_1$ and $T_2$), readout assignment and state preparation error per qubit, different subsets of qubits are not equivalent, as is shown in our results provided by the third set ($q_{27}$ - $q_{31}$), which performance is worst than the other sets.

\begin{figure}[h]
	\centering
	\includegraphics[width=0.8\linewidth]{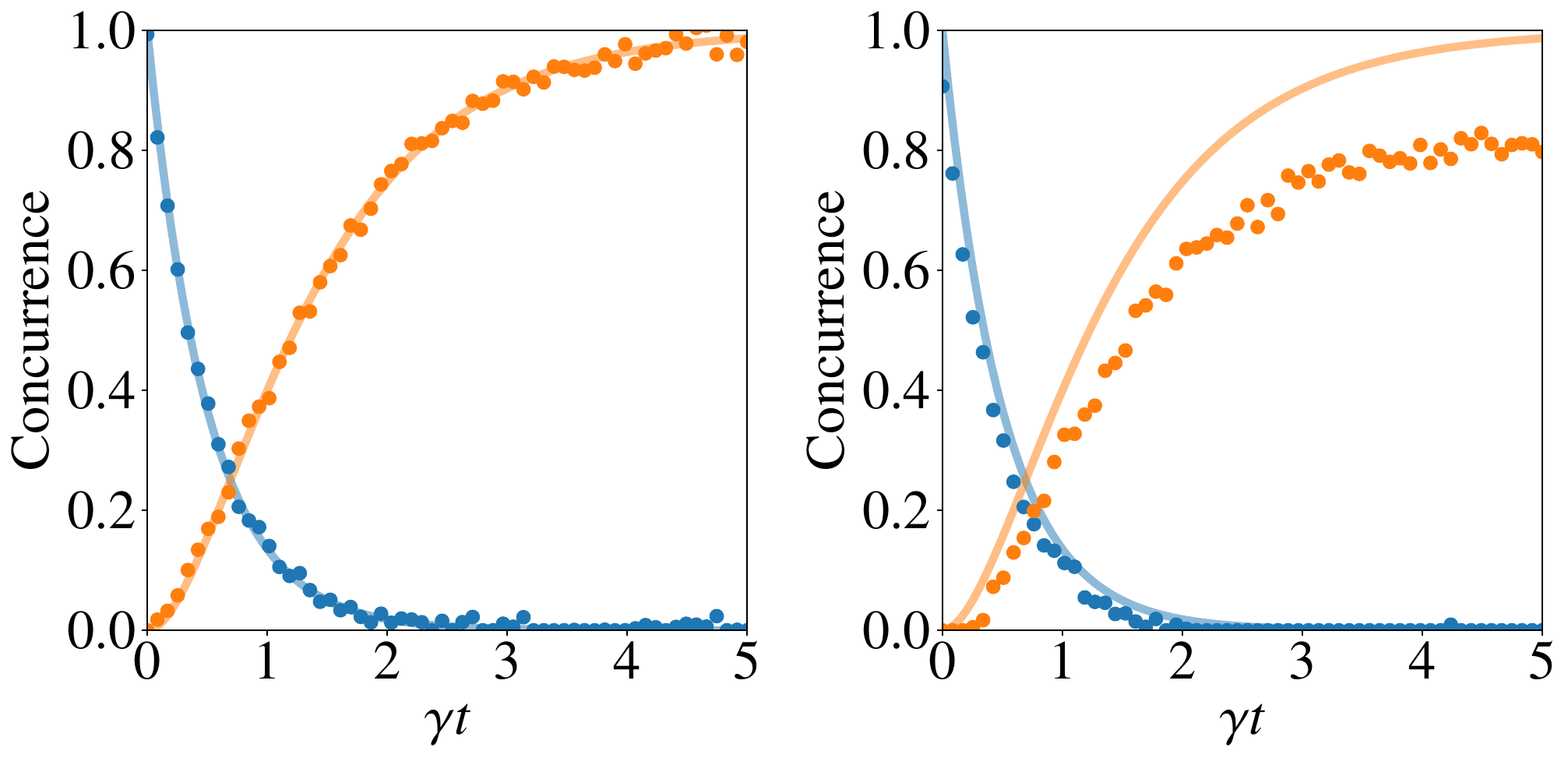}
	\caption{Concurrence as a function of $\gamma t$ for the the amplitude $\alpha =1/\sqrt{2}$ obtained using the simulator for the case without noise (left panel) and with noise considering  error mitigation (right panel). Orange dots are for environment concurrence and blue dots for system concurrence. The solid curves show the theoretical results.}
	\label{Fig08}
\end{figure}

\begin{figure}[h]
	\centering
	\includegraphics[width=0.8\linewidth]{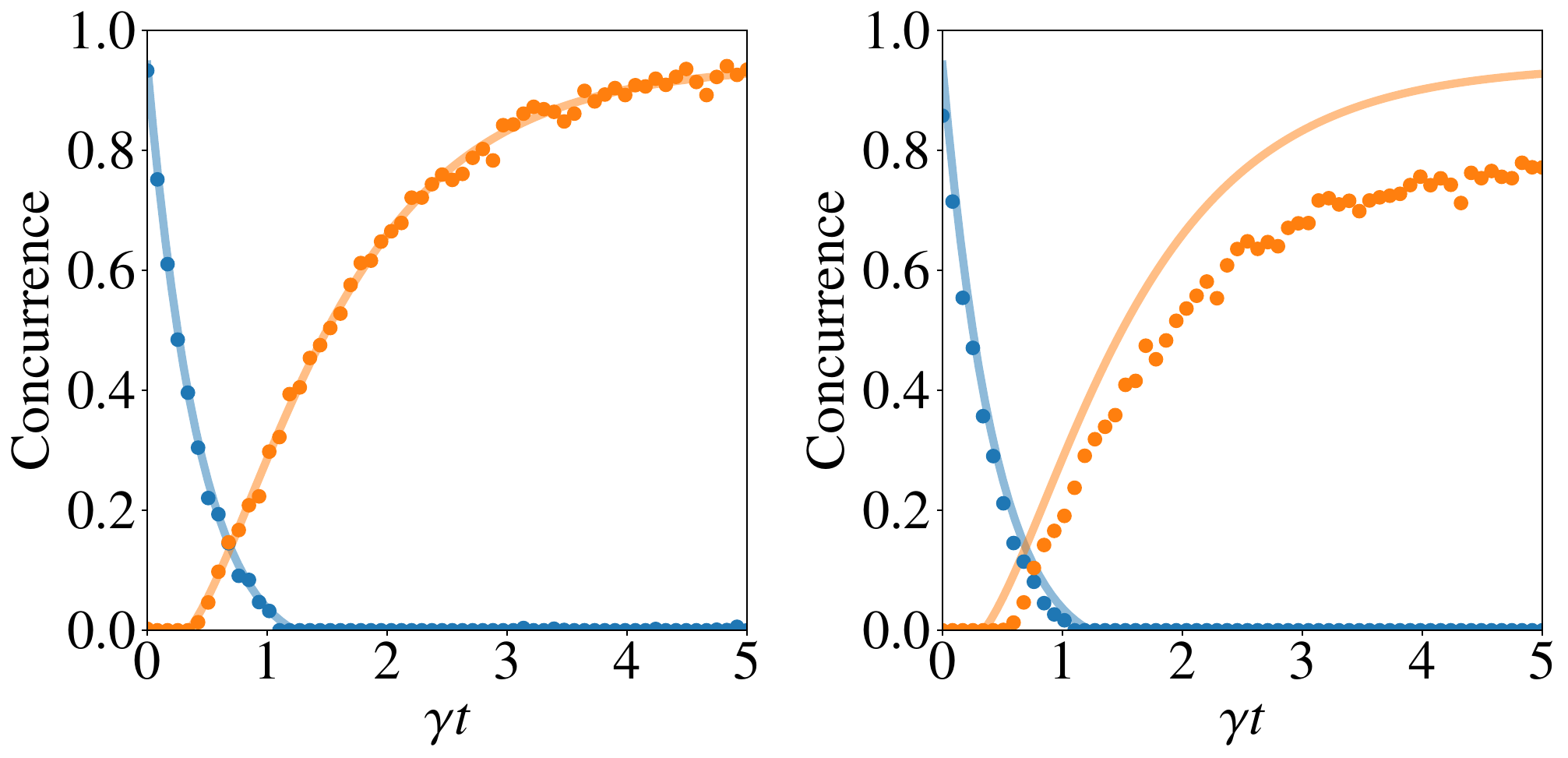}
	\caption{Concurrence as a function of $\gamma t$ for the the amplitude $\alpha =1/\sqrt{3}$ obtained using the simulator for the case without noise (left panel) and with noise considering  error mitigation (right panel). Orange dots are for environment concurrence and blue dots for system concurrence. The solid curves show the theoretical results.}
	\label{Fig09}
\end{figure}

\begin{figure}[h]
	\centering
	\includegraphics[width=0.8\linewidth]{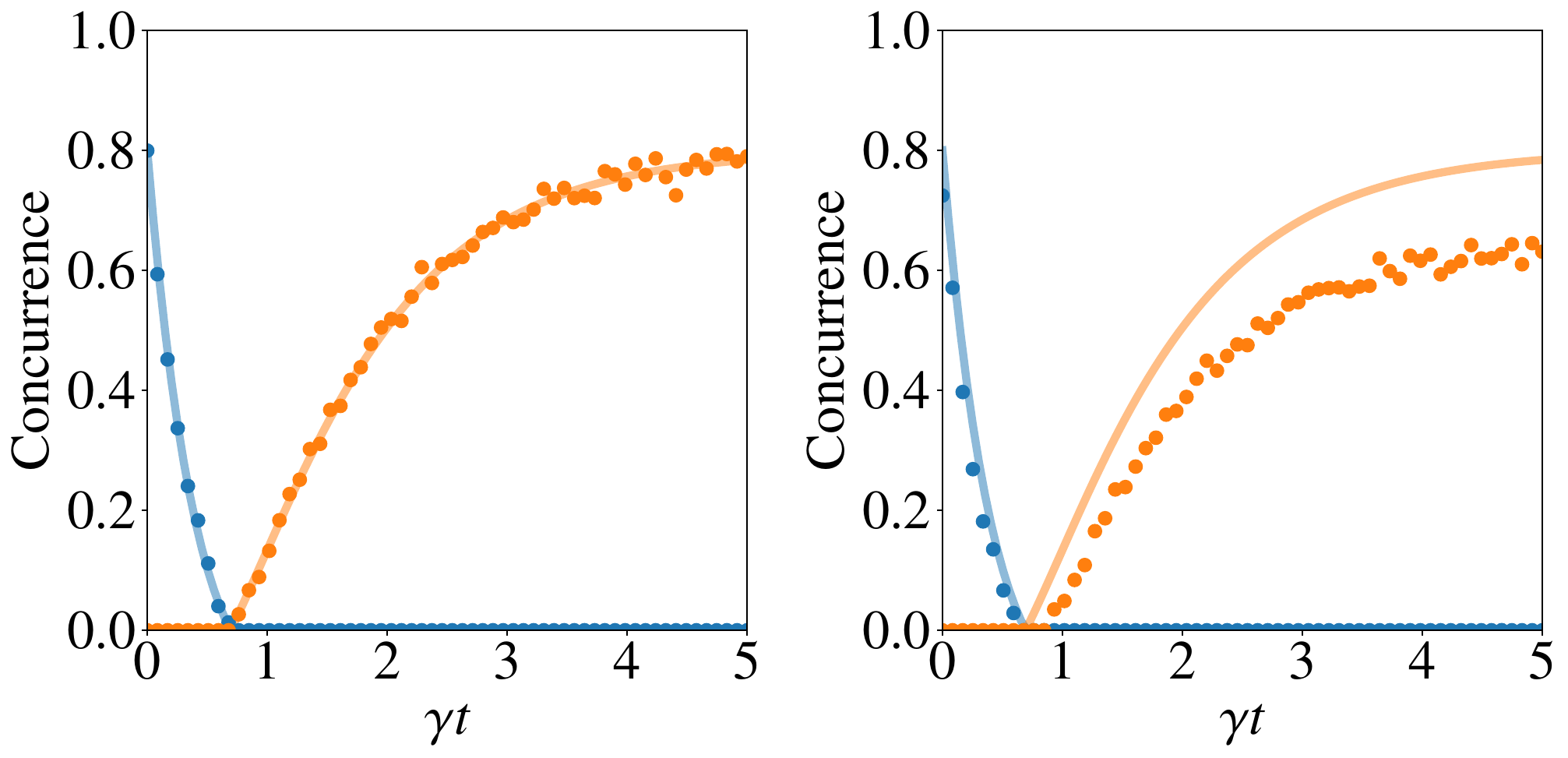}
	\caption{Concurrence as a function of $\gamma t$ for the the amplitude $\alpha =1/\sqrt{5}$ obtained using the simulator for the case without noise (left panel) and with noise considering  error mitigation (right panel). Orange dots are for environment concurrence and blue dots for system concurrence. The solid curves show the theoretical results.}
	\label{Fig10}
\end{figure}

We can get some understanding about the behavior exhibited by the entanglement estimator running the protocol for entanglement detection using the corresponding perfect simulator, and simulator with the noise model available for this backend (FakeBrisbane).  We first consider the case without the noise model, to compare with the analytical results~\cite{Lopez2008PhysRevLett}. This is shown in Fig.~\ref{Fig08}-\ref{Fig10} left panel, where we observe a good correspondence with the three cases studied. The same figures shows in the right panel, the protocole executed in the simulator with the noise model of the device and considering error mitigation and error suppression.  As we can see, there is a good agreement of Fig. \ref{Fig08}-\ref{Fig10} right panel with those corresponding to the experimental values  for each $\alpha$ carried out in set 1. 

The question arising at this point, is the entanglement estimator signaling correctly the dynamics of entanglement? To answer this question we should have at hand the real state at which the two system qubits (or the reservoir qubits) are arriving at a given time. Unfortunately this is prohibitive because in the quantum processor, a tomographic measurement is highly time demanding to be implemented.  But we could use the simulator including noise to accomplish this, which nearly describe the real experimental device.

\begin{figure}[t]
	\centering
	\includegraphics[width=0.7\linewidth]{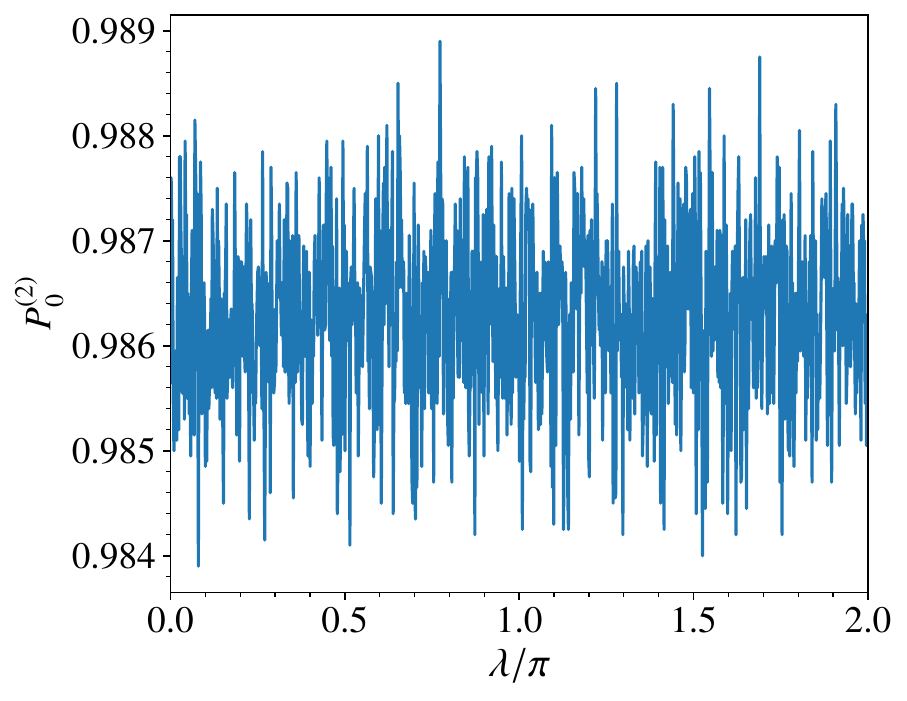}
	\caption{Population of the state $\ket{0}$ for the auxiliary qubit $q_2$, at the end of the initial preparation step (Fig.~\ref{Fig02}) using the backend FakeBrisbane.}
	\label{Fig11}
\end{figure}

To see up to what extent the initial state is modified by the noise device, we  prepare using the simulator the state for different amplitudes in terms of $\lambda/\pi$ as given in equation (\ref{Eq02}).  As the initial state preparation among qubits $q_1$ and $q_3$ involves the ancilla qubit $q_2$, any error in the gates will introduce  a deviation  from the ideal situation (ancillary final state in $|0\rangle$). As can be seen in Fig. \ref{Fig11} where the population in the state $|0\rangle$ for different amplitudes is appreciably different to one.  This deviation, originated by the imperfections in the device, is responsible for reduction of initial entanglement, as shown for $\alpha =1/\sqrt{3}$ in Fig. \ref{Fig09}, in the ideal state (left panel) as compared with the real situation (right panel). In addition this deviation from the ideal situation, is responsible for the shift in times indicating the disappearance and appearance of entanglement in the three cases shown in Fig. \ref{Fig05}-\ref{Fig07}. The sudden death times happen earlier, and sudden birth time happen later than the ideal case. 

\section{Conclusion}\label{sec4}

In this work we have implemented an experimental simulation of the entanglement dynamics of two qubits, each one being affected by an independent environment. We have designed a circuit for five qubit, and run it in \textit{ibm$\_$brisbane} , a 127 qubit quantum processor that allows to encode in linear five qubit  configuration. We used three set of qubits in this processor to encode three different entangled states. Entanglement was estimated by measuring an entanglement witness  directly related to the Wooters concurrence for the kind of initial states chosen. The entanglement measured exhibit qualitative agreement with the theoretical result both the system and environment qubits. A depletion in maximal values is produced due to a distribution of probabilities in the states of both systems arisen because of the use of the ancillary system which acquire some population, in the preparation of the initial state. This probability redistribution is superposed to the real noise of the quantum processor, that at the end cause  a shift in times $t_d$ and $t_b$ respectively that are signaling the sudden death and birth of entanglement. The present work enhance the relevance of  accesible experimental devices, such $ibm$ quantum processors,  for testing quantum mechanical predictions. It is worth to mention that in the case of pure entangled states a protocol based on two copies of the same state has been proposed~\cite{Romero2007PhysRevA} and implemented to measure concurrence in $ibmq$~\cite{Karimi2024PhysScrip}.

\section*{Acknowledgements}

We acknowledge financial support from Agencia Nacional de Investigaci\'on y Desarrollo: Financiamiento Basal para Centros Cient\'{i}ficos y Tecnol\'ogicos de Excelencia (Grant No. AFB220001), Fondecyt (grant 1231172) and Subvenci\'on a la Instalaci\'on en la Academia (grant SA77210018). Also the financial support from Universidad de Santiago de Chile: DICYT Asociativo (Grant No. 042431AA$\_$DAS). Finally, we acknowledge to students and colleagues from Physics Department at USACH, for running the program in their $ibmq$ account and provide us the corresponding data.

\newpage
\appendix

\section{Second experiment }\label{secA1}

Result for the simultaneous experiments for $\alpha=1/ \sqrt{2}$, $\alpha=1/ \sqrt{3}$ and $\alpha=1/ \sqrt{5}$, using set 2,3,1 respectively.

\begin{figure}[h]
	\centering
	\includegraphics[width=0.8\linewidth]{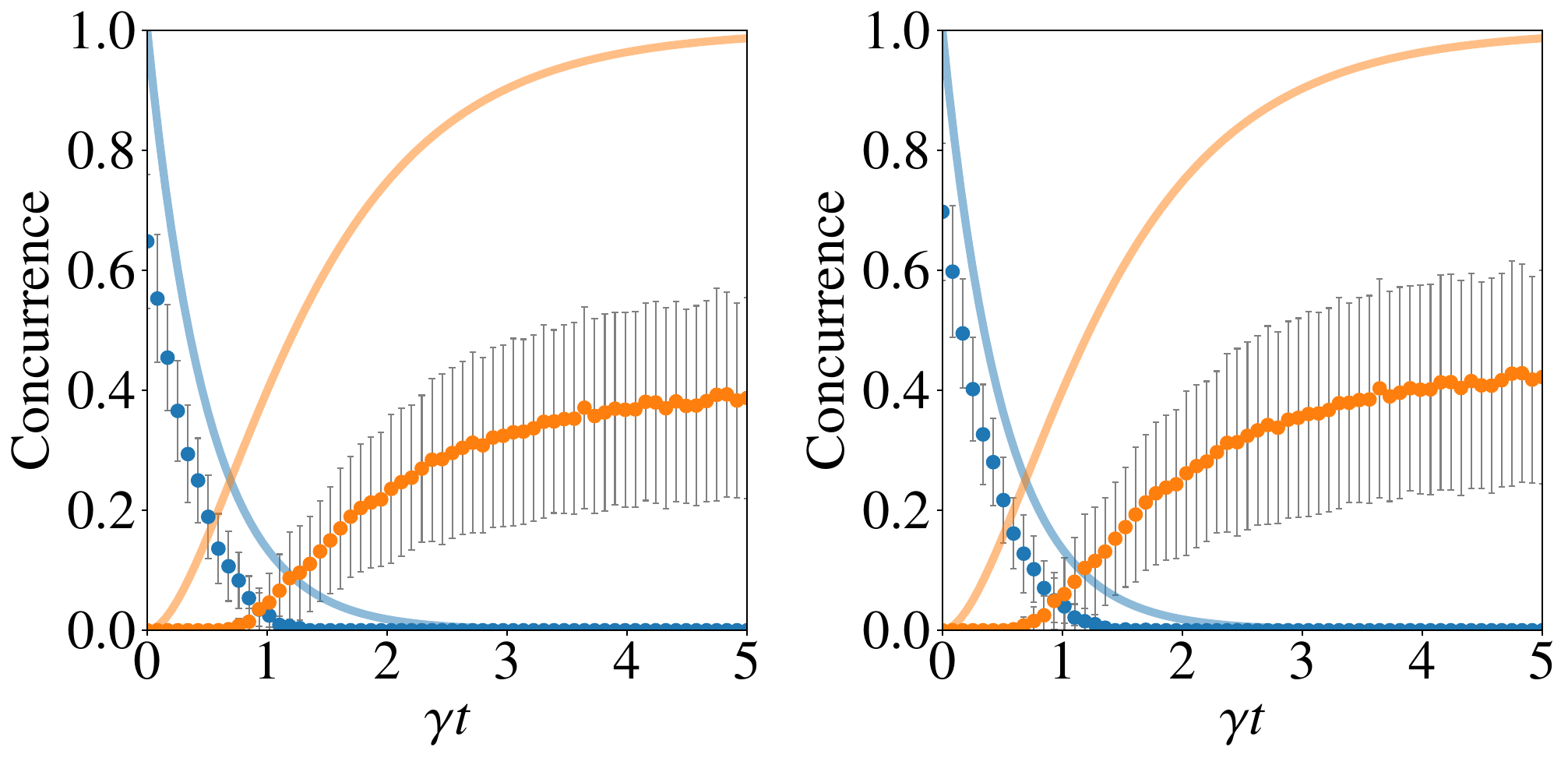}
	\caption{Concurrence as a function of $\gamma t$ for the the amplitude $\alpha =1/\sqrt{2}$ obtained using the set 2 considering  error mitigation (right panel). Orange dots are for environment concurrence and blue dots for system concurrence. The solid curves show the theoretical results.}
	\label{Fig}
\end{figure}

\begin{figure}[h]
	\centering
	\includegraphics[width=0.8\linewidth]{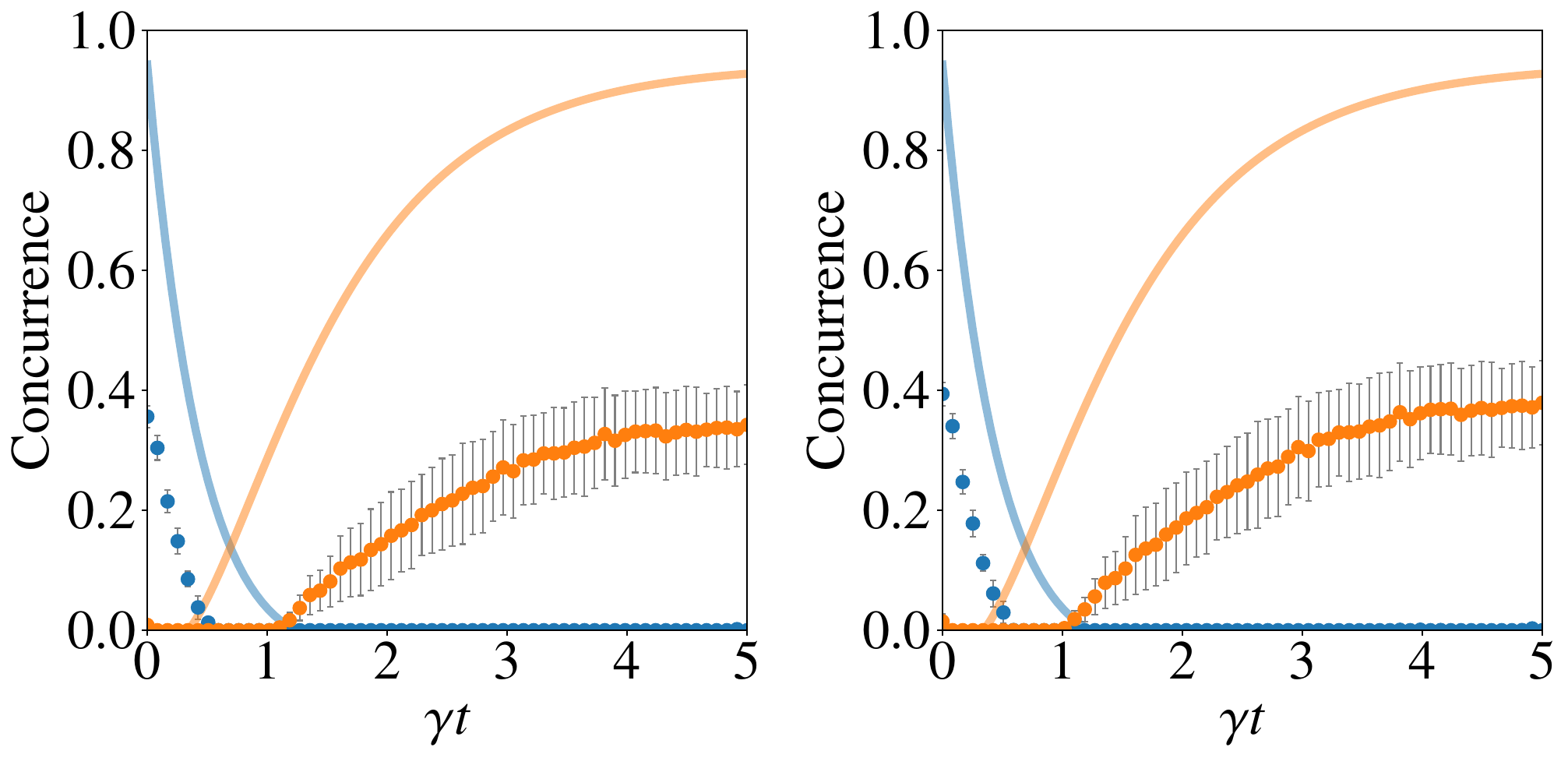}
	\caption{Concurrence as a function of $\gamma t$ for the the amplitude $\alpha =1/\sqrt{3}$ obtained using the set 3 considering  error mitigation (right panel). Orange dots are for environment concurrence and blue dots for system concurrence. The solid curves show the theoretical results.}
	\label{Fig}
\end{figure}

\begin{figure}[h]
	\centering
	\includegraphics[width=0.8\linewidth]{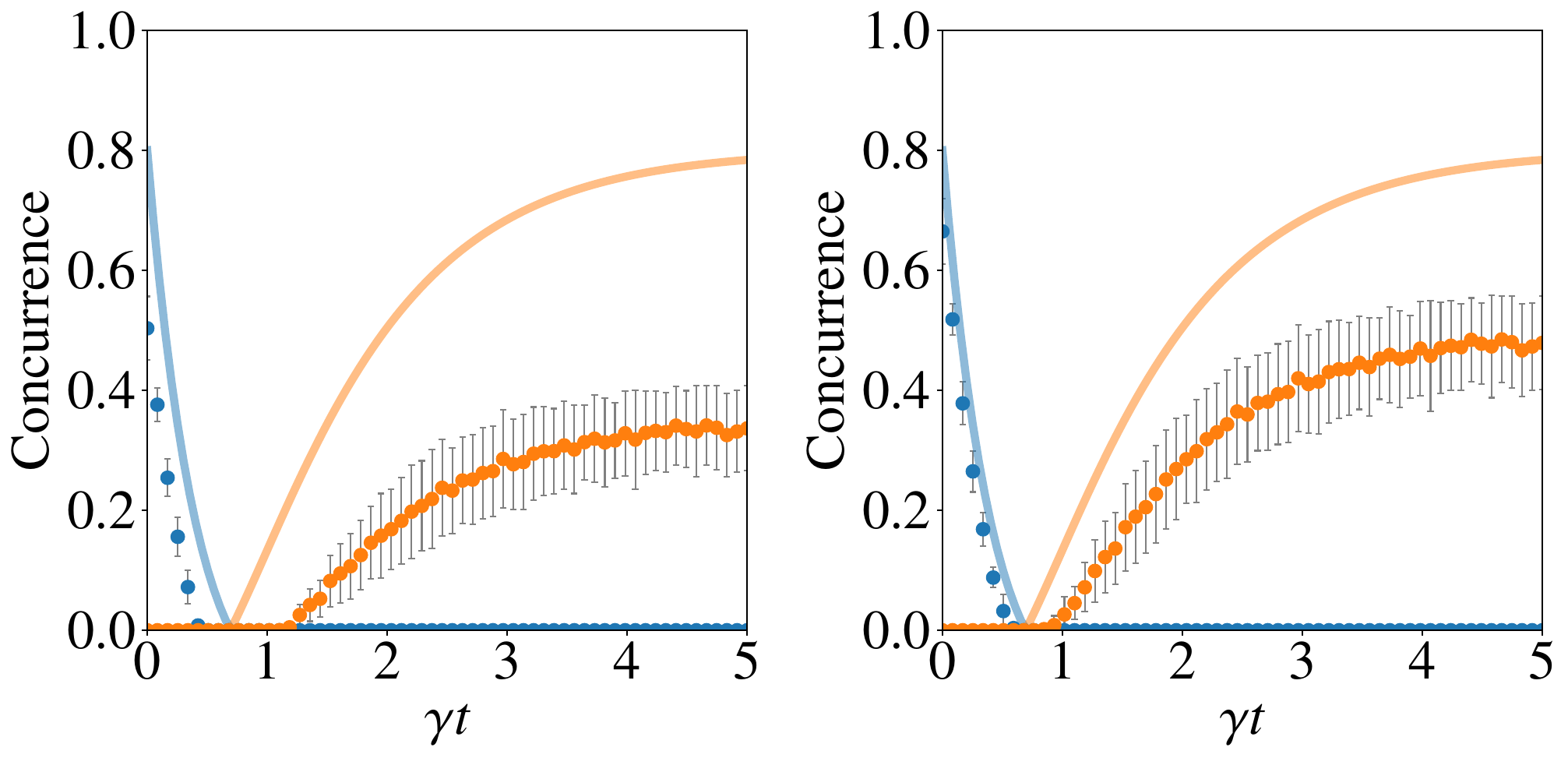}
	\caption{Concurrence as a function of $\gamma t$ for the the amplitude $\alpha =1/\sqrt{5}$ obtained using the set 1 considering  error mitigation (right panel). Orange dots are for environment concurrence and blue dots for system concurrence. The solid curves show the theoretical results.}
	\label{Fig}
\end{figure}

\newpage
\section{Third experiment}\label{secA2}

Result for the simultaneous experiments for $\alpha=1/ \sqrt{2}$, $\alpha=1/ \sqrt{3}$ and $\alpha=1/ \sqrt{5}$, using set 3,1,2 respectively.

\begin{figure}[h]
	\centering
	\includegraphics[width=0.8\linewidth]{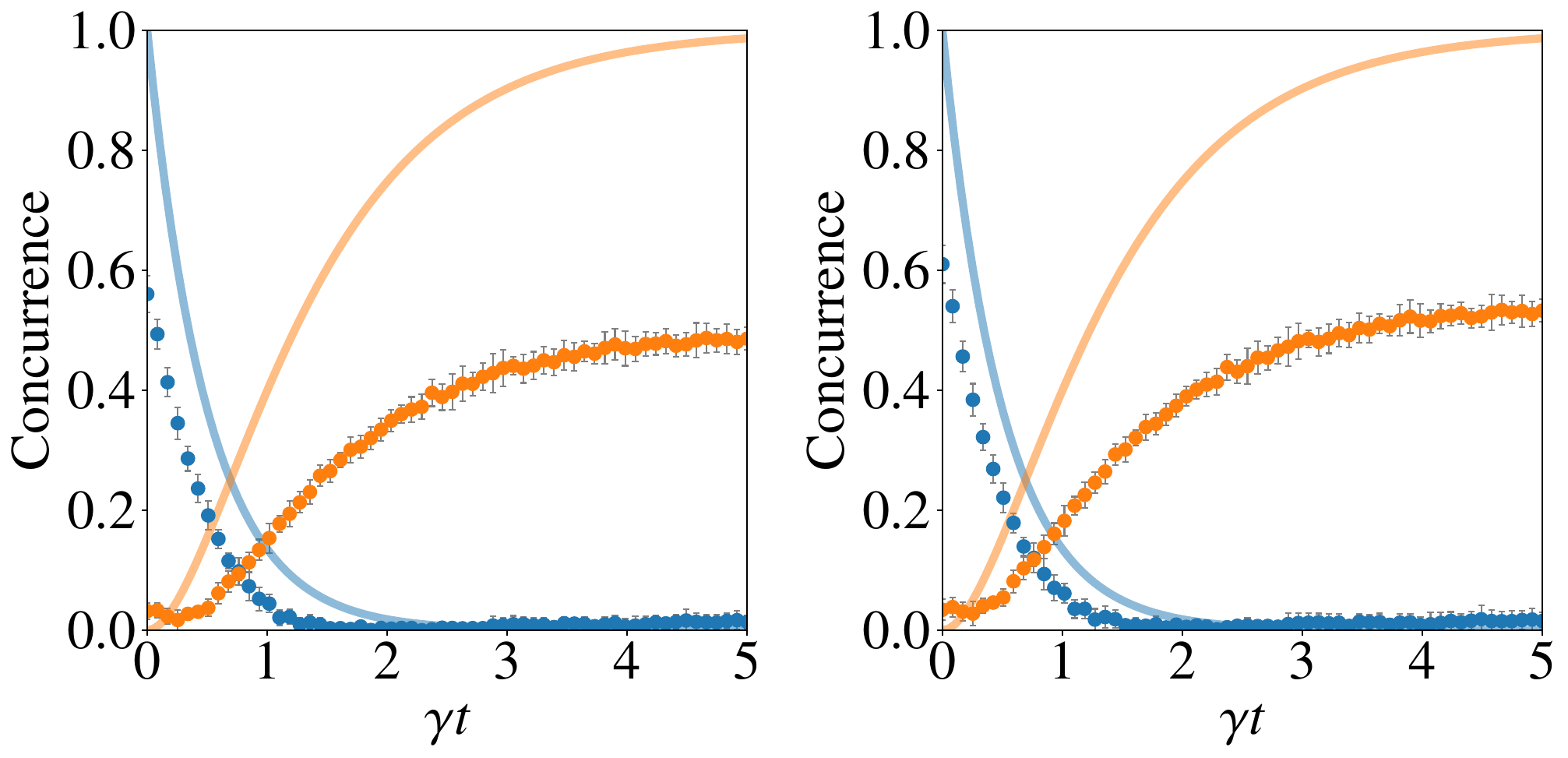}
	\caption{Concurrence as a function of $\gamma t$ for the the amplitude $\alpha =1/\sqrt{2}$ obtained using the set 2 considering  error mitigation (right panel). Orange dots are for environment concurrence and blue dots for system concurrence. The solid curves show the theoretical results.}
	\label{Fig}
\end{figure}

\begin{figure}[h]
	\centering
	\includegraphics[width=0.8\linewidth]{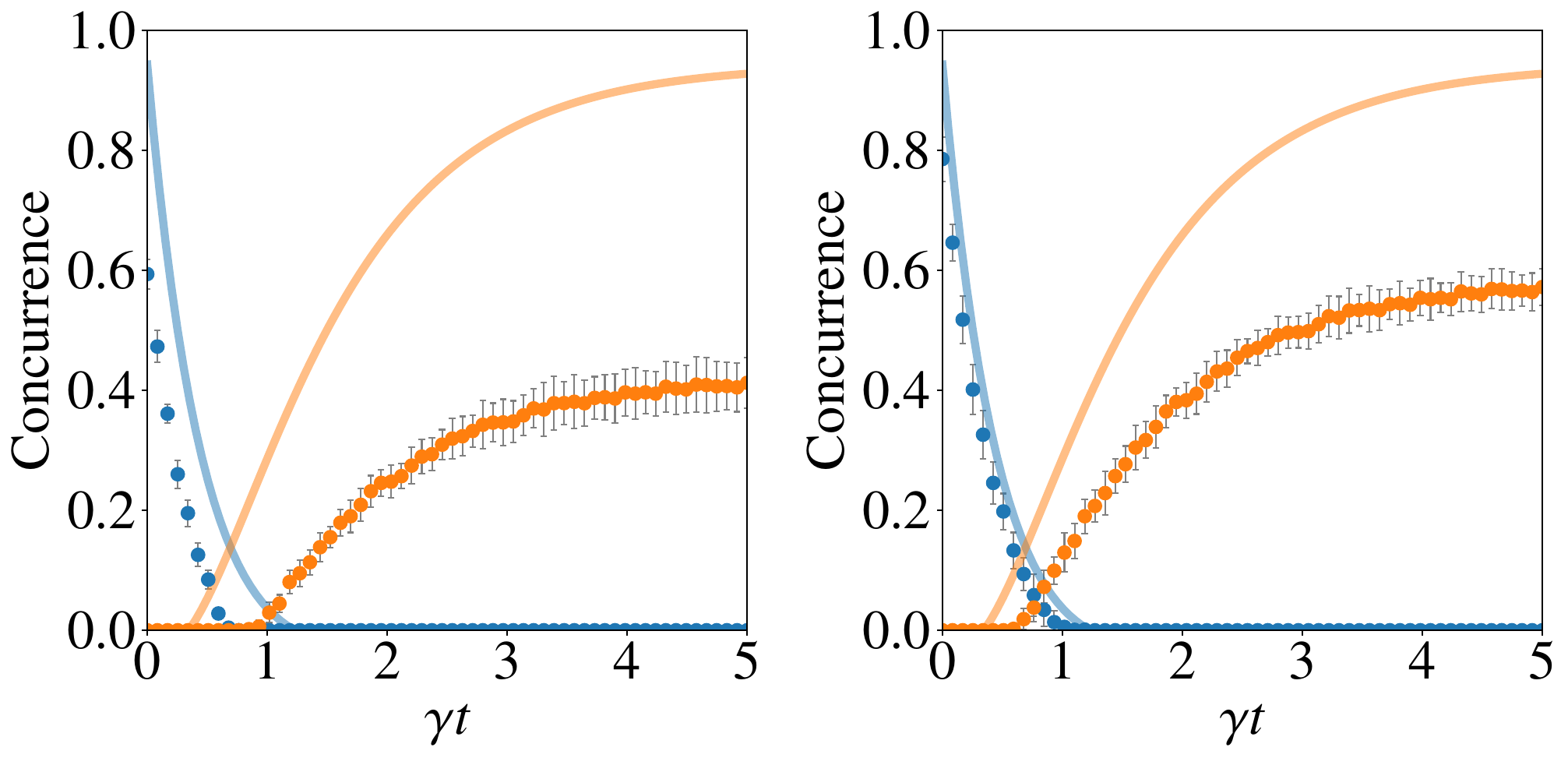}
	\caption{Concurrence as a function of $\gamma t$ for the the amplitude $\alpha =1/\sqrt{3}$ obtained using the set 3 considering  error mitigation (right panel). Orange dots are for environment concurrence and blue dots for system concurrence. The solid curves show the theoretical results.}
	\label{Fig}
\end{figure}

\begin{figure}[h]
	\centering
	\includegraphics[width=0.8\linewidth]{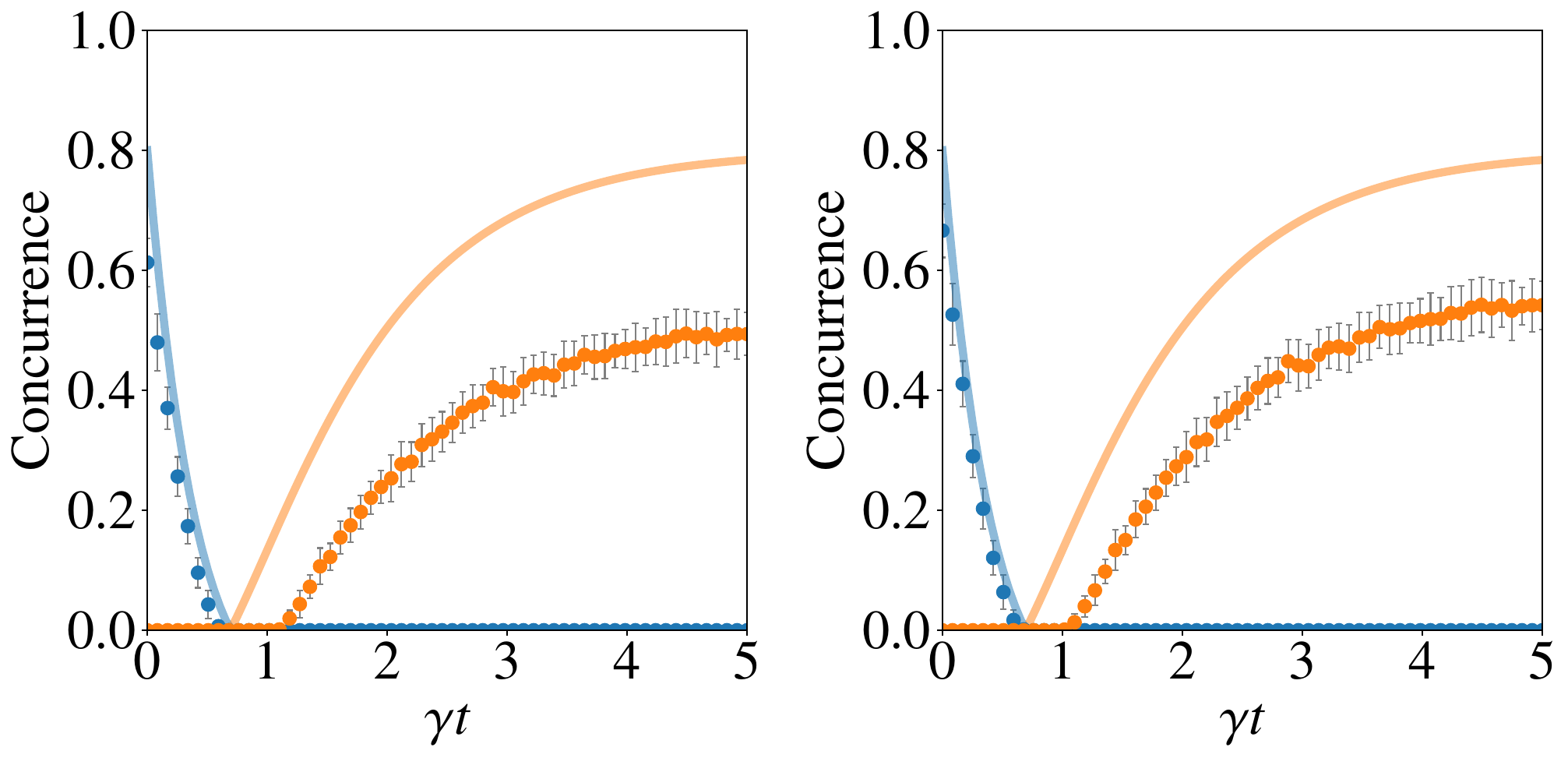}
	\caption{Concurrence as a function of $\gamma t$ for the the amplitude $\alpha =1/\sqrt{5}$ obtained using the set 1 considering  error mitigation (right panel). Orange dots are for environment concurrence and blue dots for system concurrence. The solid curves show the theoretical results.}
	\label{Fig}
\end{figure}

{}


\newpage
\begin{thebibliography}{999}
\bibitem{Bharti2022RevModPhys} K. Bharti, A. Cervera-Liera, T. Kyaw, T. Haug, S. Alperin-Lea, A. Anand, M. Degroote, H. Heimonen, J. S. Kottmann, T. Menke, W.-K. Mok, S. Sim, L.-C. Kwek, and A. Aspuru-Guzik. 
Noisy intermediate-scale quantum algorithms. 
\href{https://journals.aps.org/rmp/abstract/10.1103/RevModPhys.94.015004} 
{\textit{Rev. Mod. Phys.} \textbf{94}, 015004 (2022).}

\bibitem{Preskill2018Quantum} J. Preskill. 
Quantum Computing in the NISQ era and beyond. 
\href{https://quantum-journal.org/papers/q-2018-08-06-79/} 
{\textit{Quantum} \textbf{2}, 79 (2018).}

\bibitem{Zhang2019NatlSciRev} X. Zhang, H.-O. Li, G. Cao, M. Xiao, G.-C. Guo, and G.-P. Guo.
Semiconductor quantum computing.
\href{https://academic.oup.com/nsr/article/6/1/32/5257863}
{\textit{Natl. Sci. Rev.} \textbf{6}, 32 (2019).}

\bibitem{Loss1998PhysRevA} D. Loss and D. P. DiVincenzo.
Quantum computing with quantum dots. 
\href{https://journals.aps.org/pra/abstract/10.1103/PhysRevA.57.120} 
{\textit{Phys. Rev. A} \textbf{57}, 120 (1998).}

\bibitem{Levine2019PhysRevLett} H. Levine, A. Keesling, G. Semeghini, A. Omran, T. T. Wang, S. Ebadi, H. Bernien, M. Greiner, V. Vuleti\'c, H. Pichler, and M. D. Lukin. 
Parallel Implementation of High-Fidelity Multiqubit Gates with Neutral Atoms.
\href{https://journals.aps.org/prl/abstract/10.1103/PhysRevLett.123.170503} 
{\textit{Phys. Rev. Lett.} \textbf{123}, 170503 (2019).}

\bibitem{Henriet2020Quantum} L. Henriet, L. Beguin, A. Signoles, T. Lahaye, A. Browaeys, G.-O. Reymond, and C. Jurczak. 
Quantum computing with neutral atoms. 
\href{https://quantum-journal.org/papers/q-2020-09-21-327/} 
{\textit{Quantum} \textbf{4}, 327 (2020).}

\bibitem{Bruzewicz2019ApplPhysRev} C. D. Bruzewicz, J. Chiaverini, R. McConnell, and J. M. Sage.
Trapped-ion quantum computing: Progress and challenges.
\href{https://aip.scitation.org/doi/pdf/10.1063/1.5088164}
{\textit{Appl. Phys. Rev.} \textbf{6}, 021314 (2019).}

\bibitem{Pelucchi2021NatRevPhys}E. Pelucchi, G. Fagas, I. Aharonovich, D. Englund, E. Figueroa, Q. Gong, H. Hannes, J. Liu, C.-Y. Lu, N. Matsuda, J.-W. Pan, F. Schreck, F. Sciarrino, C. Silberhorn, J. Wang, and K. D. J\"ons.
The potential and global outlook of integrated photonics for quantum technologies. 
\href{https://www.nature.com/articles/s42254-021-00398-z} 
{\textit{Nat. Rev. Phys.} \textbf{4}, 194 (2021).}


\bibitem{Huang2020SciChinaInfSci}H.-L. Huang, D. Wu, D. Fan, and X. Zhu. 
Superconducting quantum computing: a review. 
\href{https://link.springer.com/article/10.1007/s11432-020-2881-9} 
{\textit{Sci. China Inf. Sci.} \textbf{63}, 180501 (2020).}

\bibitem{Kjaergaard2020AnnuRevCondensMatterPhys}M. Kjaergaard, M. E. Schwartz, J. Braum\"uller, P. Krantz, J. I.-J. Wang, S. Gustavsson, and W. D. Oliver. 
Superconducting Qubits: Current State of Play. 
\href{https://www.annualreviews.org/doi/abs/10.1146/annurev-conmatphys-031119-050605} 
{\textit{Annu. Rev. Condens. Matter Phys.} \textbf{11}, 369 (2020).}

\bibitem{Zhong2021PhysRevLett}H.-S. Zhong, Y.-H. Deng, J Qin, H. Wang, M.-C. Chen, L.-C. Peng, Y.-H. Luo, D. Wu, S.-Q. Gong, H. Su, Y. Hu, P. Hu, X.-Y. Yang, W.-J. Zhang, H. Li, Y. Li, X. Jiang, L. Gan, G. Yang, L. You, Z. Wang, L. Li, N.-L. Liu, J. J. Renema, C.-Y. Lu, and J.-W. Pan. 
Phase-Programmable Gaussian Boson Sampling Using Stimulated Squeezed Light. 
\href{https://journals.aps.org/prl/abstract/10.1103/PhysRevLett.127.180502} 
{\textit{Phys. Rev. Lett.} \textbf{127}, 180502 (2021).}

\bibitem{Madsen2022Nature}L. S. Madsen, F. Laudenbach, M. Falamarzi. A., F. Rortais, T. Vincent, J. F. F. Bulmer, F. M. Miatto, L. Neuhaus, L. G. Helt, M. J. Collins, A. E. Lita, T. Gerrits, S. W. Nam, V. D. Vaidya, M. Menotti, I. Dhand, Z. Vernon, N. Quesada, and J. Lavoie. 
Quantum computational advantage with a programmable photonic processor. 
\href{https://www.nature.com/articles/s41586-022-04725-x} 
{\textit{Nature} \textbf{606}, 75 (2022).}

\bibitem{Wu2021PhysRevLett}Y. Wu, W.-S. Bao, S. Cao, F. Chen, M.-C. Chen, X. Chen, T.-H. Chung, H. Deng, Y. Du, D. Fan, M. Gong, C. Guo, C. Guo, S. Guo, L. Han, L. Hong, H.-L. Huang, Y.-H. Huo, L. Li, N. Li, S. Li, Y. Li, F. Liang, C. Lin, J. Lin, H. Qian, D. Qiao, H. Rong, H. Su, L. Sun, L. Wang, S. Wang, D. Wu, Y. Xu, K. Yan, W. Yang, Y. Yang, Y. Ye, J. Yin, C. Ying, J. Yu, C. Zha, C. Zhang, H. Zhang, K. Zhang, Y. Zhang, H. Zhao, Y. Zhao, L. Zhou, Q. Zhu, C.-Y. Lu, C.-Z. Peng, X. Zhu, and J.-W. Pan. 
Strong Quantum Computational Advantage Using a Superconducting Quantum Processor. 
\href{https://journals.aps.org/prl/abstract/10.1103/PhysRevLett.127.180501} 
{\textit{Phys. Rev. Lett.} \textbf{127}, 180501 (2021).}

\bibitem{Morvan2023arXiv}A. Morvan, B. Villalonga, X. Mi, S. Mandr\`{a}, A. Bengtsson, P. V. Klimov, Z. Chen, S. Hong, C. Erickson, I. K. Drozdov, J. Chau, G. Laun, R. Movassagh, A. Asfaw, L. T.A.N. Brand\~{a}o, R. Peralta, D. Abanin, R. Acharya, R. Allen, T. I. Andersen, K. Anderson, M. Ansmann, F. Arute, K. Arya, J. Atalaya, J. C. Bardin, A. Bilmes, G. Bortoli, A. Bourassa, J. Bovaird, L. Brill, M. Broughton, B. B. Buckley, D. A. Buell, T. Burger, B. Burkett, N. Bushnell, J. Campero, H. S. Chang, B. Chiaro, D. Chik, C. Chou, J. Cogan, R. Collins, P. Conner, W. Courtney, A. L. Crook, B. Curtin, D. M. Debroy, A. Del Toro Barba, S. Demura, A. Di Paolo, A. Dunsworth, L. Faoro, E. Farhi, R. Fatemi, V. S. Ferreira, L. Flores Burgos, E. Forati, A. G. Fowler, B. Foxen, G. Garcia, E. Genois, W. Giang, C. Gidney, D. Gilboa, M. Giustina, R. Gosula, A. Grajales Dau, J. A. Gross, S. Habegger, M. C. Hamilton, M. Hansen, M. P. Harrigan, S. D. Harrington, P. Heu, M. R. Hoffmann, T. Huang, A. Huff, W. J. Huggins, L. B. Ioffe, S. V. Isakov, J. Iveland, E. Jeffrey, Z. Jiang, C. Jones, P. Juhas, D. Kafri, T. Khattar, M. Khezri, M. Kieferov\'a, S. Kim, A. Kitaev, A. R. Klots, A. N. Korotkov, F. Kostritsa, J. M. Kreikebaum, D. Landhuis, P. Laptev, K.-M. Lau , L. Laws, J. Lee, K. W. Lee, Y. D. Lensky, B. J. Lester, A. T. Lill, W. Liu, W. P. Livingston, A. Locharla, F. D. Malone, O. Martin, S. Martin, J. R. McClean, M. McEwen, K. C. Miao, A. Mieszala, S. Montazeri, W. Mruczkiewicz, O. Naaman, M. Neeley, C. Neill, A. Nersisyan, M. Newman, J. H. Ng, A. Nguyen, M. Nguyen, M. Yuezhen Niu, T. E. O'Brien, S. Omonije, A. Opremcak, A. Petukhov, R. Potter, L. P. Pryadko, C. Quintana, D. M. Rhodes, E. Rosenberg, C. Rocque, P. Roushan, N. C. Rubin, N. Saei, D. Sank, K. Sankaragomathi, K. J. Satzinger, H. F. Schurkus, C. Schuster, M. J. Shearn, A. Shorter, N. Shutty, V. Shvarts, V. Sivak, J. Skruzny, W. C. Smith, R. D. Somma, G. Sterling, D. Strain, M. Szalay, D. Thor, A. Torres, G. Vidal, C. Vollgraff Heidweiller, T. White, B. W. K. Woo, C. Xing, Z. J. Yao, P. Yeh, J. Yoo, G. Young, A. Zalcman, Y. Zhang, N. Zhu, N. Zobrist, E. G. Rieffel, R. Biswas, R. Babbush, D. Bacon, J. Hilton, E. Lucero, H. Neven, A. Megrant, J. Kelly, I. Aleiner, V. Smelyanskiy, K. Kechedzhi, Y. Chen, S. Boixo.
Phase transition in Random Circuit Sampling. 
\href{https://arxiv.org/abs/2304.11119} 
{arXiv:2304.11119 \textbf{[quant-ph]} 2023.}

\bibitem{Fowler2012PhysRevA}A. G. Fowler, M. Mariantoni, J. M. Martinis, and A. N. Cleland. 
Surface codes: Towards practical large-scale quantum computation. 
\href{https://journals.aps.org/pra/abstract/10.1103/PhysRevA.86.032324} 
{\textit{Phys. Rev. A} \textbf{86}, 032324 (2012).}

\bibitem{Daley2022Nature}A. J. Daley, I. Bloch, C. Kokail, S. Flannigan, N. Pearson, M. Troyer, and P. Zoller. 
Practical quantum advantage in quantum simulation. 
\href{https://www.nature.com/articles/s41586-022-04940-6} 
{\textit{Nature} \textbf{607}, 667 (2022).}

\bibitem{IBMQ} Official link: \href{https://quantum-computing.ibm.com}{www.quantum-computing.ibm.com}

\bibitem{Horodecki2009RevModPhys}R. Horodecki, P. Horodecki, M. Horodecki, and K. Horodecki.
Quantum entanglement.
\href{https://journals.aps.org/rmp/abstract/10.1103/RevModPhys.81.865}
{\textit{Rev. Mod. Phys.} \textbf{81}, 865 (2009).}

\bibitem{wooters1998}W. K. Wootters. 
Entanglement of Formation of an Arbitrary State of Two Qubits.
\href{https://journals.aps.org/prl/abstract/10.1103/PhysRevLett.80.2245}
{\textit{Phys. Rev. Lett.} \textbf{80}, 2245 (1998).}

\bibitem{eberly2004}T. Yu and J. H. Eberly. 
Finite-Time Disentanglement Via Spontaneous Emission.
\href{https://journals.aps.org/prl/abstract/10.1103/PhysRevLett.93.140404}
{\textit{Phys. Rev. Lett.} \textbf{93}, 140404 (2004).}

\bibitem{Lopez2008PhysRevLett}C. E. L\'opez, G. Romero, F. Lastra, E. Solano, and J. C. Retamal.
Sudden Birth versus Sudden Death of Entanglement in Multipartite Systems.  
\href{https://journals.aps.org/prl/abstract/10.1103/PhysRevLett.101.080503}
{\textit{Phys. Rev. Lett.} \textbf{101}, 080503 (2008).}

\bibitem{davidovich1}M. P. Almeida, F. de Melo, M. Hor-Meyll, A. Salles, S. P. Walborn, P. H. S. Ribeiro, and L. Davidovich.
Environment-Induced Sudden Death of Entanglement
\href{https://www.science.org/doi/abs/10.1126/science.1139892}
{\textit{Science} \textbf{316}, 579 (2007).}

\bibitem{davidovich2}G. H. Aguilar, A. Vald\'es-Hern\'andez, L. Davidovich, S. P. Walborn, and P. H. S. Ribeiro.
Experimental Entanglement Redistribution under Decoherence Channels
\href{https://journals.aps.org/prl/abstract/10.1103/PhysRevLett.113.240501}
{\textit{Phys. Rev. Lett.} \textbf{113}, 240501 (2014).}

\bibitem{davidovich3}O. J. Far\'{i}as, G. H. Aguilar, A. Vald\'es-Hern\'andez, P. H. S. Ribeiro, L. Davidovich, and S. P. Walborn.
Observation of the Emergence of Multipartite Entanglement Between a Bipartite System and its Environment.
\href{https://journals.aps.org/prl/abstract/10.1103/PhysRevLett.109.150403}
{\textit{Phys. Rev. Lett.} \textbf{109}, 150403 (2012).}

\bibitem{GarciaPerez2020NpjQuantumInf}G. Garc\'ia-P\'erez, M. A. C. Rossi, and S. Maniscalco.
IBM Q Experience as a versatile experimental testbed for simulating open quantum systems.  
\href{https://www.nature.com/articles/s41534-019-0235-y}
{\textit{Npj Quantum Inf.} \textbf{6}, 1 (2020).}

\bibitem{Santos2006PhysRevA}M. F. Santos, P. Milman, L. Davidovich, and N. Zagury.
Direct measurement of finite-time disentanglement induced by a reservoir.  
\href{https://journals.aps.org/pra/abstract/10.1103/PhysRevA.73.040305}
{\textit{Phys. Rev. A} \textbf{73}, 040305(R) (2006).}

\bibitem{PaulDNation2021PRXQuantum}Paul D. Nation, Hwajung Kang, Neereja Sundaresan, and Jay M. Gambeta.
Scalable Mitigation of Measurement Errors on Quantum Computers.  
\href{https://journals.aps.org/prxquantum/abstract/10.1103/PRXQuantum.2.040326}
{\textit{PRX Quantum} \textbf{2}, 040326 (2021).}

\bibitem{Romero2007PhysRevA} G. Romero, C. E. L\'opez, F. Lastra, E. Solano, and J. C. Retamal. Direct measurement of concurrence for atomic two-qubit pure states. \href{https://journals.aps.org/pra/abstract/10.1103/PhysRevA.75.032303}{\textit{Phys. Rev. A} \textbf{75}, 032303}

\bibitem{Karimi2024PhysScrip} N. Karimi, S. N. Elyasi, and M. Yahyavi. Implementation and measurement of quantum entanglement using IBM quantum platforms. \href{https://iopscience.iop.org/article/10.1088/1402-4896/ad3518}{\textit{Phys. Scr.} \textbf{99}, 045121 (2024).}




\end{thebibliography}
\end{document}